\documentclass[prd,superscriptaddress,unsortedaddress,twocolumn,showpacs,floatfix]{revtex4}    

\usepackage{epsfig}
\usepackage{dcolumn}

\newcommand{\KLpienu}{\mbox{$K_L\to \pi^{\pm} e^{\mp} \nu$}}
\newcommand{\KLpimunu}{\mbox{$K_L\to \pi^{\pm} \mu^{\mp} \nu$}}
\newcommand{\Kethree}{\mbox{$K_{e3}$}}
\newcommand{\IKe}{\mbox{$I_K^e$}}
\newcommand{\vus}{\mbox{$|V_{us}|$}}

\newcommand{\MK}{\mbox{$M_{K}$}}

\newcommand{\mpi}{\mbox{$M_{\pi}$}}
\newcommand{\sqmpi}{\mbox{$M_{\pi}^2$}}
\newcommand{\sqsqmpi}{\mbox{$M_{\pi}^4$}}

\newcommand{\SQMV}{\mbox{$M_{V}^2$}}

\newcommand{\fhat}{\hat{f}_{+}(t)}
\newcommand{\fhatlin}{\hat{f}^{lin}_{+}(t)}
\newcommand{\fhatquad}{\hat{f}^{quad}_{+}(t)}
\newcommand{\fhatpole}{\hat{f}^{pole}_{+}(t)}

\newcommand{\fhatz}{\hat{f}_{+}^z(t)}
\newcommand{\Ft}{{F}_{+}^z(t)}
\newcommand{\Fzero}{{F}_{+}^z(0)}
\newcommand{\Ftzero}{{F}_{+}^z(t_0)}

\newcommand{\dGdt}{d\Gamma/dt}

\newcommand{\lplus}{\mbox{$\lambda_{+}$}}
\newcommand{\lplusp}{\mbox{$\lambda_{+}'$}}
\newcommand{\lpluspp}{\mbox{$\lambda_{+}''$}}

\newcommand{\az}{\mbox{$a_{1}$}}
\newcommand{\azz}{\mbox{$a_{2}$}}
\newcommand{\azzz}{\mbox{$a_{3}$}}

\newcommand{\topt}{0.39}  
\newcommand{\ASUM}{13}     
\newcommand{\ASQSUM}{170}  

\newcommand{\azval}{1.024}
\newcommand{\azerrstat}{0.028}
\newcommand{\azerrsyst}{0.027}   

\newcommand{\azchisq}{62.5/66}

\newcommand{\ikval}{0.15406}     
\newcommand{\ikerrstat}{0.00020}
\newcommand{\ikerrsyst}{0.00019}
\newcommand{\ikerrthry}{0.0023}  
\newcommand{\ikerrtot}{0.0023}

\newcommand{\AZVAL}{1.023}
\newcommand{\AZERRSTAT}{0.028}
\newcommand{\AZERRSYST}{0.029}   
\newcommand{\AZERRTOT}{0.040}

\newcommand{\AZZVAL}{0.75}
\newcommand{\AZZERRSTAT}{1.58}
\newcommand{\AZZERRSYST}{1.47}   
\newcommand{\AZZERRTOT}{2.16}

\newcommand{\AZCHISQ}{62.3/65}
\newcommand{\AZCOR}{-0.064}

\newcommand{\IKVAL}{0.15392}
\newcommand{\IKERRSTAT}{0.00035}
\newcommand{\IKERRSYST}{0.00033}
\newcommand{\IKERRTHRY}{0.00006}
\newcommand{\IKERRTOT}{0.00048}

\begin{document}


\title{Improved $\KLpienu$ Form Factor and Phase Space Integral \\
        with Reduced Model Uncertainty}

\newcommand{\UAz}{University of Arizona, Tucson, Arizona 85721}
\newcommand{\UCLA}{University of California at Los Angeles, Los Angeles,
                    California 90095} 
\newcommand{\UCSD}{University of California at San Diego, La Jolla,
                   California 92093} 
\newcommand{\EFI}{The Enrico Fermi Institute, The University of Chicago, 
                  Chicago, Illinois 60637}
\newcommand{\UB}{University of Colorado, Boulder, Colorado 80309}
\newcommand{\ELM}{Elmhurst College, Elmhurst, Illinois 60126}
\newcommand{\FNAL}{Fermi National Accelerator Laboratory, 
                   Batavia, Illinois 60510}
\newcommand{\Osaka}{Osaka University, Toyonaka, Osaka 560-0043 Japan} 
\newcommand{\Rice}{Rice University, Houston, Texas 77005}
\newcommand{\UVa}{The Department of Physics and Institute of Nuclear and 
                  Particle Physics, University of Virginia, 
                  Charlottesville, Virginia 22901}
\newcommand{\UW}{University of Wisconsin, Madison, Wisconsin 53706}

\affiliation{\UAz}
\affiliation{\UCLA}
\affiliation{\UCSD}
\affiliation{\EFI}
\affiliation{\UB}
\affiliation{\ELM}
\affiliation{\FNAL}
\affiliation{\Osaka}
\affiliation{\Rice}
\affiliation{\UVa}
\affiliation{\UW}

\author{E.~Abouzaid}	  \affiliation{\EFI}
\author{M.~Arenton}       \affiliation{\UVa}
\author{A.R.~Barker}      \affiliation{\UB}
\author{L.~Bellantoni}    \affiliation{\FNAL}
\author{A.~Bellavance}    \affiliation{\Rice}
\author{E.~Blucher}       \affiliation{\EFI}
\author{G.J.~Bock}        \affiliation{\FNAL}
\author{E.~Cheu}          \affiliation{\UAz}
\author{R.~Coleman}       \affiliation{\FNAL}
\author{M.D.~Corcoran}    \affiliation{\Rice}
\author{B.~Cox}           \affiliation{\UVa}
\author{A.R.~Erwin}       \affiliation{\UW}
\author{A.~Glazov}        \affiliation{\EFI}
\author{A.~Golossanov}    \affiliation{\UVa}
\author{Y.B.~Hsiung}      \affiliation{\FNAL}
\author{H.~Huang}         \affiliation{\UB}
\author{D.A.~Jensen}      \affiliation{\FNAL}
\author{R.~Kessler}       \affiliation{\EFI}
\author{H.G.E.~Kobrak}    \affiliation{\UCSD}
\author{K.~Kotera}        \affiliation{\Osaka}
\author{A.~Ledovskoy}     \affiliation{\UVa}
\author{P.L.~McBride}     \affiliation{\FNAL}

\author{E.~Monnier}
   \altaffiliation[Permanent address ]{C.P.P. Marseille/C.N.R.S., France}
   \affiliation{\EFI}

\author{H.~Nguyen}       \affiliation{\FNAL}
\author{R.~Niclasen}     \affiliation{\UB} 
\author{E.J.~Ramberg}    \affiliation{\FNAL}
\author{R.E.~Ray}        \affiliation{\FNAL}
\author{M.~Ronquest}	 \affiliation{\UVa}
\author{J.~Shields}      \affiliation{\UVa}
\author{W.~Slater}       \affiliation{\UCLA}
\author{D.~Smith}	 \affiliation{\UVa}
\author{N.~Solomey}      \affiliation{\EFI}
\author{E.C.~Swallow}    \affiliation{\EFI}\affiliation{\ELM}
\author{P.A.~Toale}      \affiliation{\UB}
\author{R.~Tschirhart}   \affiliation{\FNAL}
\author{Y.W.~Wah}        \affiliation{\EFI}
\author{J.~Wang}         \affiliation{\UAz}
\author{H.B.~White}      \affiliation{\FNAL}
\author{J.~Whitmore}     \affiliation{\FNAL}
\author{M.~Wilking}      \affiliation{\UB}
\author{B.~Winstein}     \affiliation{\EFI}
\author{R.~Winston}      \affiliation{\EFI}
\author{E.T.~Worcester}  \affiliation{\EFI}
\author{T.~Yamanaka}     \affiliation{\Osaka}
\author{E.~D.~Zimmerman} \affiliation{\UB}

\collaboration{The KTeV Collaboration}

\date{\today}

\begin{abstract}
  Using the published KTeV sample of 2~million 
  $\KLpienu$ decays \cite{ktev:kl3ff} 
  and a new form factor expansion with a rigorous bound on higher 
  order terms \cite{Hill:Ke3zff06},
  we present a new determination of the $\KLpienu$ form factor
  and phase space integral.
  Compared to the previous KTeV result \cite{ktev:kl3ff},
  the uncertainty in the new form factor expansion is
  negligible and results in an overall uncertainty
  in the phase space integral ($\IKe$) that is a factor of two smaller:
  $\IKe = \IKVAL \pm \IKERRTOT$.
\end{abstract}



\maketitle

\begin{figure}
\centering
\psfig{figure=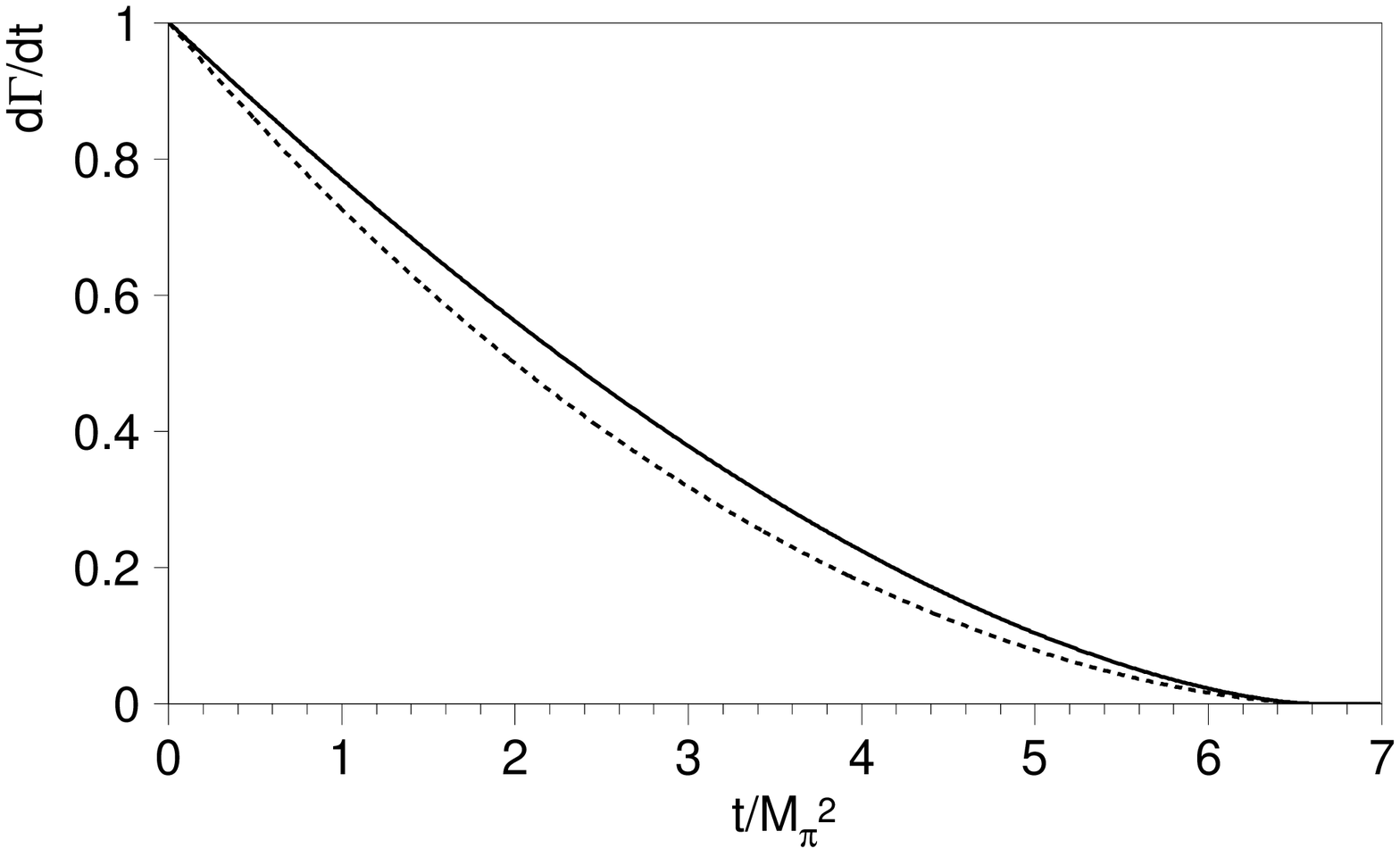,width=\linewidth}
\vspace{-0.5cm}
\caption{
  calculated $t$-distribution, $\dGdt$, for $\KLpienu$
  phase space (dashed), and for phase space modified
  by the form factor in \cite{ktev:kl3ff} (solid line).
        }
\label{fig:phase}
\end{figure}

Our previous analysis of semileptonic form factors \cite{ktev:kl3ff}
contributed to improving the precision in the 
Cabibbo-Kobayashi-Maskawa (CKM) matrix element $\vus$ 
\cite{Cabibbo63,KM73,LR84}.
The published form factor results were based on
linear, quadratic and pole models:
\begin{eqnarray}
   \fhatlin  & = & 1 + \lplus  \frac{t}{\sqmpi}  \label{eq:lin} \\
   \fhatquad & = & 1 + \lplusp \frac{t}{\sqmpi} 
                 + \lpluspp \frac{1}{2}\frac{t^2}{\sqsqmpi}
   \label{eq:quad} \\
   \fhatpole & = & \frac{\SQMV}{\SQMV - t} \label{eq:pole}
\end{eqnarray}
where 
$t = (P_K - p_{\pi})^2$ is the expansion 
parameter as a function of the kaon and pion four-momenta,
$\fhat \equiv f_{+}(t)/f_{+}(0)$ is the normalized form factor,
and $\lplus$, $\lplusp$, $\lpluspp$, $\SQMV$
are the form factor parameters measured in a fit to data.
Figure~\ref{fig:phase} shows the $t$-distribution for 
$\KLpienu$ phase space with and without a form factor.
The best fit to the KTeV data, 
with $\chi^2/dof = 62/65$,
\footnote{In \cite{ktev:kl3ff}, 
   the reported number of degrees of freedom (dof) 
   is off by 1; the correct dof values are reported here.}
was obtained with $\fhatquad$ 
(Eq.~\ref{eq:quad}).
The pole model, with one free parameter,
also gives a good fit with $\chi^2/dof = 66/66$.
The linear model, with $\chi^2/dof = 81/66$, is disfavored.
Although the data are well described by the 
quadratic and pole models,
the corresponding phase space integrals ($\IKe$) differ
by 0.7\%, which is much larger than the experimental precision
of 0.3\%. This 0.7\% difference was arbitrarily included in the 
$\IKe$ uncertainty for the determination of $\vus$.

The form factor models in Eqs.~\ref{eq:lin}-\ref{eq:pole}
are theoretically flawed:
$\fhatpole$ makes the unlikely assumption that there
is only one pole,
and truncated Taylor expansions violate unitarity.
These issues are addressed in a new parametrization
based on analyticity and bounds
from $\tau \to K\pi\nu$ decays \cite{Hill:Ke3zff06}.
In this paper, we present an updated form factor and phase space integral
based on our published data and this new parametrization.

The new parametrization expands in the variable
\begin{equation}
  z(t,t_0) \equiv
 \frac{\sqrt{t_{+} - t} - \sqrt{t_{+} - t_0}}
      {\sqrt{t_{+} - t} + \sqrt{t_{+} - t_0}}~,
\end{equation}
where $t_{+} = (\MK + \mpi)^2$, and $t_0$ is an arbitrary
anchor point. Note that our previous expansion parameter,
$t/\sqmpi$, has a range of values from zero to about six; 
the new expansion parameter is limited to $|z| < 0.2$.
The form factor expansion in this new variable is
\begin{equation}
   \Ft = \Ftzero\frac{\phi(t_0,t_0,Q^2)}{\phi(t,t_0,Q^2)} 
                 \sum_{k=0}^{\infty} a_k(t_0,Q^2) z(t,t_0)^k ~,
   \label{eq:fz}
\end{equation}
where $a_0 \equiv 1$,
$a_k$~($k\ge 1$) are coefficients to be determined, 
$Q^2$ is the invariant momentum-squared of the current in the 
operator product expansion,
and the function $\phi$ is defined as
\begin{eqnarray}\label{eq:phiplus}
&&\phi(t,t_0,Q^2) = \sqrt{1\over 32\pi} {z(t,0)\over -t}
\left(z(t,-Q^2)\over -Q^2-t\right)^{3/2} \nonumber \\
&& \times
\left(z(t,t_0)\over t_0-t\right)^{-1/2}
\left(z(t,t_-)\over t_--t\right)^{-3/4}
{t_+-t\over (t_+-t_0)^{1/4}}  
\end{eqnarray}
with $t_{\pm} = (\MK \pm \mpi)^2$.
The advantage of this expansion is that higher
order terms are bounded by~\cite{Hill:Ke3zff06}
\begin{equation}
   A^2 \equiv \sum_{k=0}^{\infty} {a_k^2}  \le \ASQSUM ~.
   \label{eq:bound}
\end{equation}

Using the previously published KTeV sample of 2~million $\KLpienu$ decays, 
the smallest overall error on
$\IKe$ is obtained from a second order ``quadratic-$z$''
fit to $\Ft$ in Eq.~\ref{eq:fz}.
The expansion coefficients
$\az(t_0,Q^2)$  and $\azz(t_0,Q^2)$ 
depend on the choice of $t_0$ and $Q^2$ ;
however, the function $\Ft$, the calculated
$\IKe$ value, and the uncertainty on $\IKe$
are independent of $t_0,Q^2$ \footnote{The assertion that
  $\Ft$, $\IKe$ and $\delta(\IKe)$ are independent of
  $t_0$ and $Q^2$ has been verified by varying $t_0,Q^2$
  over a large range of values.}.
We choose to fit with $Q^2=2$~GeV$^2$/c$^2$ and  $t_0 = \topt t_{-}$
because this results in no correlation between $\az$ and $\azz$ 
if the experimental acceptance of
$\KLpienu$ decays is uniform as a function of $t$.

We fit the KTeV $\KLpienu$ sample to the normalized form factor,
\begin{equation}
  \fhatz = \Ft / \Fzero ~.
\end{equation}
The results are
\begin{eqnarray}
  {\az}  & = &  \AZVAL  \pm \AZERRSTAT(stat) \pm \AZERRSYST(syst) \\
         & = &  \AZVAL  \pm \AZERRTOT  \\
  {\azz} & = &  \AZZVAL \pm \AZZERRSTAT(stat) \pm \AZZERRSYST(syst) \\
         & = &  \AZZVAL \pm \AZZERRTOT  \\
  \chi^2/dof & = & \AZCHISQ  \\
  \rho_{12}  & = & \AZCOR ~.
\end{eqnarray}
The second-order term ${\azz}$ is consistent with zero,
and the small correlation ($\rho_{12} =  \AZCOR$)
is due to the slightly non-uniform acceptance
of the KTeV detector \footnote{The KTeV detector acceptance
  for $\KLpienu$ decays varies by less than 8\% over the 
  entire range of $t$.}.
The systematic uncertainties are obtained by scaling the ratio of
systematic-to-statistical uncertainties for the quadratic-$t$
model in Table~I of \cite{ktev:kl3ff}, 
\begin{equation}
   \sigma_{syst}(\az,\azz) = \sigma_{stat}(\az,\azz) \times
   \frac{\sigma_{syst}(\lplusp,\lpluspp)}{\sigma_{stat}(\lplusp,\lpluspp)}~.
\end{equation}

Setting $a_k=0$ for $k \ge 3$,
the resulting phase space integral is
\begin{eqnarray}
   \IKe & = & \IKVAL   \pm \IKERRSTAT(stat) 
                       \pm \IKERRSYST(syst)  \nonumber \\
        &   &   ~~~~~~ \pm \IKERRTHRY(th)              \\
        & = & \IKVAL \pm \IKERRTOT \label{eq:ikresult}~,
\end{eqnarray}
where the uncertainties are statistical, experimental systematic,
and theoretical modeling of the form factor.
The theory error of $\pm \IKERRTHRY(th)$ is estimated
by substituting the maximum allowed value for $\azzz$
as follows,
\begin{eqnarray}
   \azzz & = & \sqrt{A_{+}^2 - {a_0}^2 - min(\az^2) - min(\azz^2) } \\
         & \simeq & \sqrt{\ASQSUM - 1 - 1 - 0 }  \sim \ASUM ~,
\end{eqnarray}
and then repeating the fit for $\az$ and $\azz$.
Note that a linear fit in $z$ gives a statistical error
on $\IKe$ that is almost a factor of two smaller than the error
from the second-order 
fit \footnote{
  Results from a linear fit to Eq.~\ref{eq:fz} are: 
  $\az = \azval \pm \azerrstat(stat) \pm \azerrsyst(syst)$, 
  $\chi^2/dof = \azchisq$, 
  $\IKe = \ikval \pm \ikerrstat(stat) \pm \ikerrsyst(syst) 
                 \pm \ikerrthry(th) = 
          \ikval \pm \ikerrtot$.}.
However, the linear fit relies
on $\delta \azz \sim \ASUM$ from the bound in Eq.~\ref{eq:bound},
and the corresponding uncertainty in $\IKe$ is $\sim 0.0023$,
more than four times larger than the overall error from the 
second order fit (Eq.~\ref{eq:ikresult}).

Comparing to our previous results in \cite{ktev:kl3ff},
our new central value for $\IKe$ lies between the
$\IKe$ values determined from the quadratic-$t$ 
and pole models.
The new experimental error (statistical plus systematic), 
$\delta(\IKe) = \IKERRTOT$, is slightly larger than the 
experimental error reported in \cite{ktev:kl3ff}
because both $\KLpienu$ and $\KLpimunu$ were used to
determine the form factors in \cite{ktev:kl3ff},
while only the $\Kethree$ mode is considered in this
updated result.
The overall error of $\delta\IKe = \IKERRTOT$ (Eq.~\ref{eq:ikresult})
is two times smaller than our previously reported error;
this reduced error is achieved by reducing the form factor 
model uncertainty from $\pm 0.00095$ in \cite{ktev:kl3ff} 
to a negligible $\IKERRTHRY$.

\begin{figure}[hb]
\centering
\psfig{figure=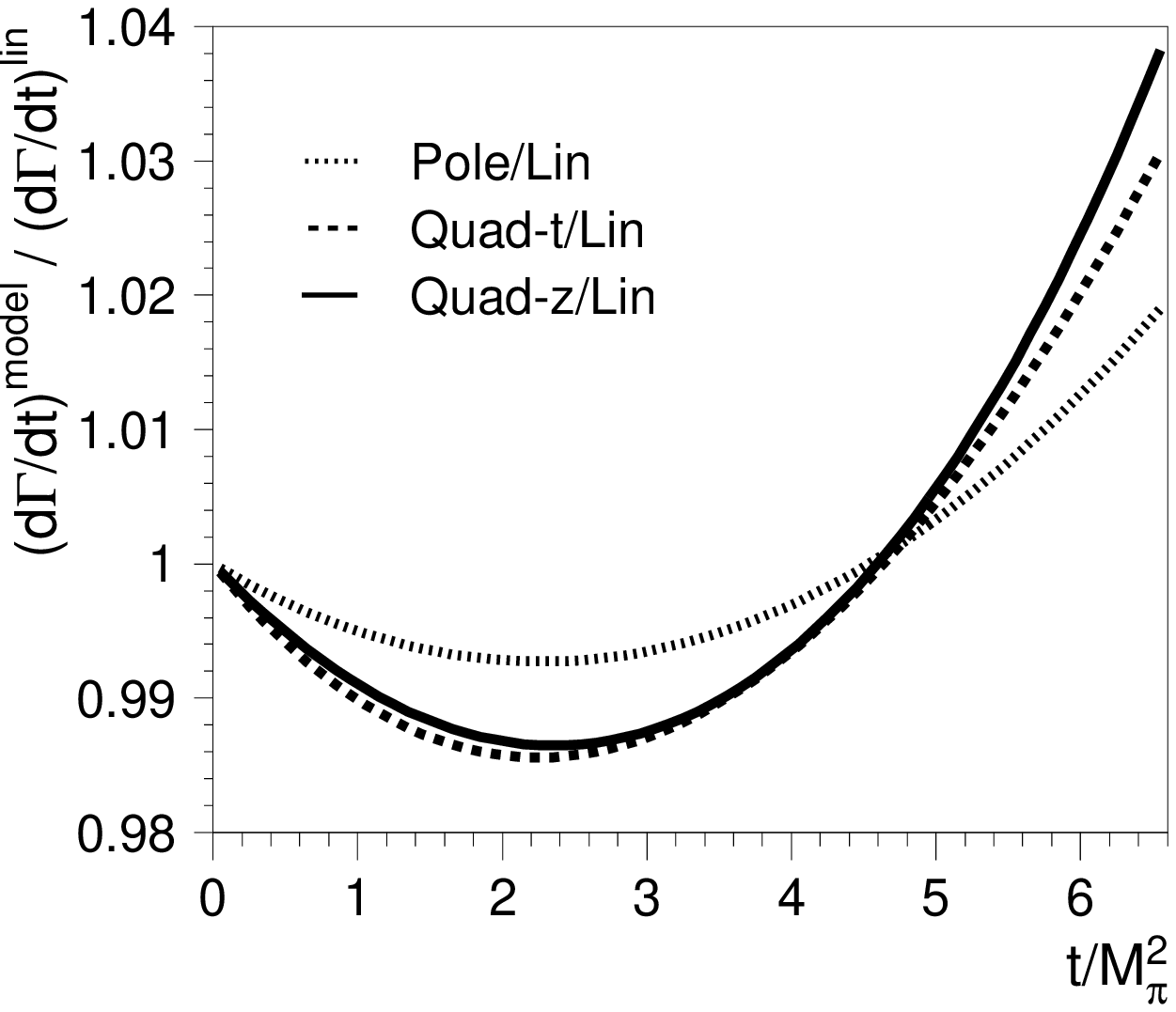,width=\linewidth}
\vspace{-0.7cm}
\caption{
  Calculated phase space ratios versus $t/\sqmpi$,
  $(\dGdt)^{model}/(\dGdt)^{lin}$,
  where $model=$ pole (Eq.~\ref{eq:pole}), 
  quadratic-$t$ (Eq.~\ref{eq:quad})
  and quadratic-$z$ (Eq.~\ref{eq:fz}).
  The reference model 
  $(\dGdt)^{lin}$ uses $\fhatlin = 1 + 0.0283t/\sqmpi$.
        }
\label{fig:ratios}
\end{figure}

Figure~\ref{fig:ratios} shows the $\dGdt$ distribution 
for different models used to fit the KTeV data. 
In all cases the distribution is divided by $(\dGdt)^{lin}$
from the linear model (Eq.~\ref{eq:lin} with $\lplus = 0.0283$) 
so that deviations from the 
linear model are more clearly visible. 
Our quadratic-$z$ fit (Eq.~\ref{eq:fz}) is much closer
to our previous best result with a quadratic-$t$ model
(Eq.~\ref{eq:quad}).  
A simple parametrization of the $t$-dependence for our
fit quadratic-$z$ model is
\begin{eqnarray}
  (\dGdt)^{\rm quad-z} & = & 
      (\dGdt)^{lin} \times [ 1 - (0.01094)\hat{t} + \nonumber \\
    (0.0020286)\hat{t}^2 & + & (0.78398\times 10^{-4})\hat{t}^3]~,
\end{eqnarray}
where $\hat{t} = t/\sqmpi$.

We gratefully acknowledge the support and effort of the Fermilab
staff and the technical staffs of the participating institutions for
their vital contributions.  This work was supported in part by the U.S. 
Department of Energy, The National Science Foundation and The Ministry of
Education and Science of Japan. 
We also wish to thank Richard Hill for useful discussions
regarding semileptonic form factors.

\bibliography{ke3zff}

\end{document}